\documentclass[%
nofootinbib,
 amsmath,amssymb,
 aps, physrev,
prd,
]{revtex4-2}

\usepackage{graphicx}
\usepackage{dcolumn}
\usepackage{bm}

\usepackage[colorlinks,linkcolor=magenta,citecolor=magenta,urlcolor=magenta]{hyperref}

\usepackage{bm,graphicx,dcolumn,epstopdf,epsf, latexsym,mathbbol, amssymb,amsmath,color,slashed, mathrsfs,mathcomp,simplewick}
\usepackage[center]{subfigure}
\usepackage{multirow}
\usepackage{graphicx}
\usepackage{makecell}
\usepackage{epstopdf}
\usepackage[table]{xcolor}
\newcommand{\orcid}[1]{\href{https://orcid.org/#1}{\includegraphics[width=10pt]{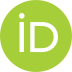}}}
\usepackage{booktabs} 
\graphicspath{{figs/}} 

\begin{document}


\title{Power Law Plateau Inflation and Primary Gravitational Waves in the light of ACT}

\author{Abolhassan Mohammadi \orcid{0000-0003-1228-9107}}
\email{abolhassnm@zjut.edu.cn; gmail.com}
\author{Yogesh \orcid{0000-0002-7638-3082}}%
 \email{yogesh@zjut.edu.cn, yogeshjjmi@gmail.com}
 \affiliation{Institute of Theoretical Physics \& Cosmology, School of Science,
Zhejiang University of Technology, Hangzhou, China. }

\author{Anzhong Wang\orcid{0000-0002-8852-9966}}
\email{Anzhong$_$Wang@baylor.edu}
\affiliation{GCAP-CASPER,  Department of Physics and Astronomy, Baylor University, Waco, TX 76798-7316, USA}


\begin{abstract}
We investigate Power-Law Plateau (PLP) inflation in standard gravity and its consistency with ACT DR6 data. While many inflationary models, including the Starobinsky inflation, are disfavored by ACT observations, the PLP potential remains viable across a broad range of its parameters. 
Then, the dynamics of the reheating phase are investigated, 
where we mainly focus on the reheating temperature and its relationship with the inflationary phase and primordial gravitational waves. Incorporating the overproduction of the primordial gravitational waves can affect the effective number of relativistic species during the bounce. The constraint data on $\Delta N_{\rm eff}$ can impose a lower bound on the reheating temperature. This constraint will be more efficient for a stiff equation of state. It is determined that for $\omega_{re} > 0.58$, this constraint would be efficient. Combining the result of the reheating temperature and the inflationary phase, it is concluded that to have both a viable result standing in $1\sigma$ of ACT DR6 and also to satisfy the reheating lower bound, the total number of e-folds during the inflationary phase should be $N_k \lesssim 62$. Higher e-folds of expansion result in a reheating temperature below the bound, which is disfavored. Finally, for the constraint values of the reheating temperature, the energy spectrum of the gravitational waves has been explored. The results indicate that there is a higher chance of detection for lower reheating temperatures and higher reheating equation of state. 
\end{abstract}

\maketitle


\section{Introduction \label{introduction}}
Over the past decades, the scenario of inflation has become the cornerstone of modern cosmology in resulting several theoretical issues of the Big Bang Model. The concept of inflation originates from the works of \cite{starobinsky:1980te,Guth:1980zm,Albrecht:1982wi,Linde:1981mu,Linde:1983gd}, and it has since been widely accepted and modified in various ways. There are different proposals for the scalar field that drives inflation, such as non-canonical inflation \cite{Barenboim:2007ii,Franche:2010yj,Unnikrishnan:2012zu,Gwyn:2012ey,Rezazadeh:2014fwa,Cespedes:2015jga,Stein:2016jja,Pinhero:2017lni}, tachyon inflation \cite{Fairbairn:2002yp,Mukohyama:2002cn,Feinstein:2002aj,Padmanabhan:2002cp}, DBI inflation \cite{Spalinski:2007dv,Bessada:2009pe,Weller:2011ey,Nazavari:2016yaa,Amani:2018ueu,Mohammadi:2018zkf}. The scenario has also been considered in modified gravity theories~\cite{Yogesh:2025wak,Pozdeeva:2020apf,Guo:2010jr,Gialamas:2023flv,Satoh:2008ck,Jiang:2013gza,Koh:2014bka,Bernardo:2025lie,Ahmed:2025rrg,Wolf:2025ecy} and quantum gravity \cite{Agullo:2013ai,Zhu:2017jew,Li:2018opr,Li:2018fco,Li:2019ipm,Mohammadi:2024bye}. In addition, different mechanisms of inflation have been introduced such as warm inflation \cite{berera1995warm,berera2000warm,
BasteroGil:2004tg,
Rosa:2018iff,Bastero-Gil:2019gao,Sayar:2017pam,Akhtari:2017mxc,Sheikhahmadi:2019gzs,Bastero-Gil:2017wwl,Chakraborty:2025jof}, constant-roll inflation \cite{Motohashi:2014ppa,Odintsov:2017hbk,Oikonomou:2017bjx,Mohammadi:2019qeu,Mohammadi:2020ftb,Mohammadi:2022vru}. Besides its theoretical variety, the scenario has received strong observational support, including WMAP \cite{WMAP:2010qai}, Planck \cite{Akrami:2018odb}, and the most recent one, Atacama Cosmology Telescope (ACT) \cite{ACT:2025fju, ACT:2025tim}. By the updating observational data, some inflationary models has been disfavored 
including some popular inflationary models, such as Starobinsky inflation.
Recently a number of works has been carried out to test various models against the latest observational data
~\cite{Kallosh:2025rni,Kallosh:2025ijd,Pallis:2025gii,Gao:2025viy,Okada:2025lpl,McDonald:2025tfp,Bianchi:2025tyl,Odintsov:2025wai,Kohri:2025lau,Chakraborty:2025oyj,Pallis:2025nrv,Frolovsky:2025iao,Yi:2025dms,Maity:2025czp,Byrnes:2025kit,Zharov:2025evb}.

ACT DR6 data predict a larger scalar spectral index, with $n_s = 0.9743 \pm 0.0034$, which puts popular inflationary models under pressure. Due to this new observation, it is necessary to reconsider  inflationary models and explore their consistency with the new data. Here, we will consider the Power-Law Plateau kind of potentials, which has been motivated by the supersymmetry \cite{Dimopoulos:2014boa,Dimopoulos:2016zhy,Yogesh:2024vcl}, as $V(\phi) = V_0 \left( {\phi^n \over \phi^n + m^n} \right)^q$, where $n$ and $q$ are real free parameters. This class of potentials, 
provides a flexible framework for exploring inflationary scenario. To preserve the plateau shape of the potential, it is assumed that the mass scale $m$ is smaller than the field $\phi$; otherwise, it would be indistinguishable from the monomial potential. Our consideration 
to be presented in this paper shall show that the result of the model perfectly agrees with the ACT DR6 for a wide range of the two free parameters $n$ and $q$: they all stand in the $1\sigma$ region of data. 

At the end of the inflationary phase, the universe is cold and empty. Therefore, to have a smooth transition to the radiation-dominant phase and recover the big-bang cosmology, the universe is required to warm up and be filled with particles. This occurs in the reheating phase right following the inflation. In this reheating phase, the energy of inflation decays into the standard particles. The produced particles interact each other and thermalize the universe. The reheating phase is usually parameterized by three parameters: the number of e-folds during this phase, $N_{re}$, its reheating temperature $T_{re}$ and effective equation of state, $\omega_{re}$. The  reheating temperature should be above the BBN temperature, $T_{BBN} = 4 \; {\rm MeV}$ \cite{Kawasaki:1999na,Kawasaki:2000en,Hasegawa:2019jsa} but smaller than GUT temperature, $T_{GUT} \simeq 10^{16} {\rm GeV}$. However, the former bound does not include the effects of the primordial gravitational waves (PGWs) \cite{Boyle:2005se,Watanabe:2006qe,Saikawa:2018rcs,Caprini:2018mtu,Figueroa:2019paj,Bernal:2019lpc,Bernal:2020ywq,Bhattacharya:2020lhc}. 
The PGWs, which are generated by the tensor fluctuations during the inflationary phase, are scale-invariant, and their amplitude is estimated by the Hubble scale. Unlike the scalar perturbations, the amplitude of PGWs is not known, although an upper bound exists and is often expressed in terms of the tensor-to-scalar ratio. These perturbations re-enter the horizon after the inflation and evolve through the reheating, radiation-dominant,  matter-dominant and dark-energy dominated phases, so that the shape and strength of the spectral density of PGWs are sensitive to the energy scale of inflation and the history of the universe. During the reheating phase, the modes in the range $k_{re} < k < k_e$ re-enter the horizon, which affect the spectral energy density of the PGWs that is measured today \cite{Bernal:2019lpc,Maity:2024odg,Maity:2024cpq,Ghoshal:2024gai}. Consequently, due to the similar behavior of the energy density of the GW to radiation, they impact the relativistic degree of freedom to radiation, indicated by $\Delta N_{\rm eff}$. In general, this effect on the relativistic degree of freedom is efficient for $\omega > 1/3$. The CMB observation for this extra degree of freedom stats $\Delta N_{\rm eff} \geq 0.17$ based on the measurement obtained from ACT with Planck datasets \cite{ACT:2025fju, ACT:2025tim}. It has been demonstrated that this constraint on $\Delta N_{\rm eff}$ can result in a lower bound on the reheating temperature. 

In this work, we show that the PLP potential agrees with the ACT DR6 data, and the resulting $r-n_s$ stands in the $1\sigma$ region for a wide range of the parameters $n$ and $q$. Then, using the result obtained for the inflationary sector regarding the values of the Hubble parameters at the horizon crossing and the end of inflation, we determine the lower bound on the temperature. It is realized that this lower bound is efficient for $\omega_{re} \geq 0.58$, and for lower $\omega_{re}$ would be smaller than $T_{BBN}$. Considering the reheating phase, one finds that to agree with the data from the inflationary phase and also to satisfy the reheating temperature bound, the total number of e-folds during inflation should be $N_k \lesssim 62$. 

This paper is organized as follows: In Sec.\ref{model}, we introduce the model briefly and outline the main quantities of the inflationary phase. Next, we study the dynamics of inflation and discuss the reheating temperature, as well as its relationship to the inflationary quantities. Then, we obtain a lower bound on the reheating temperature based on the effective number of relativistic degrees of freedom. In Sec.\ref{results}, we first discuss the consistency of the model with ACT data, then discuss the result of the new bound on the reheating temperature, and finally, we examine the energy spectrum of the gravitational waves. In Sec.\ref{conclusion}, we summarize our main results.

\section{The model \label{model}}
Considering a scalar field  minimally coupled to gravity, the action of the model could be written as
\begin{equation}\label{action}
    S = \int d^x \; \sqrt{-g} \; \left( \frac{M_p^2}{2} \; R - \frac{1}{2} g^{\mu\nu} \nabla_\mu \phi \nabla_\nu \phi - V(\phi) \right), 
\end{equation}
where $g$ is the determinant of the metric $g_{\mu\nu}$, $R$ is the Ricci scalar built from the metric, $\phi$ is the scalar field which drives inflation, and $V(\phi)$ is the potential of the scalar field. Also, $M_p$ is the reduced Planck mass defined as $M_p^2 = (8 \pi G)^{-1}$. Taking the variation of the action and utilizing a spatially flat FLRW metric leads to the following dynamical equations 
\begin{eqnarray}\label{friedmann}
    H^2 & = & \frac{1}{3 M_p^2} \; \left( \frac{1}{2} \; \dot\phi^2 + V(\phi) \right), \nonumber \\
    \dot{H} & = &- \frac{1}{2 M_p^2} \; \dot\phi^2,
\end{eqnarray}
and the field equation of motion  reads as
\begin{equation}\label{field_eom}
    \ddot\phi + 3 H \dot\phi + V'(\phi) = 0,
\end{equation}
in which a prime indicates the derivative with respect to the scalar field. During inflation, it is assumed that the scalar field is at the top of its potential and rolls down slowly. Due to these slowly rolling, the potential remains almost constant, leading to an almost-constant Hubble parameter, whereby a quasi-de Sitter phase of expansion is achieved. The slow-rolling behavior is described by the slow-roll parameters, which must be small during the inflationary phase. These parameters are defined as
\begin{equation}\label{srp}
    \epsilon = \frac{M_p^2}{2} \; \frac{V'^2}{V^2}, \qquad 
    \eta = M_p^2 \frac{V''}{V}. 
\end{equation}
Smallness of $\epsilon \ll 1$ indicates that the kinetic term is much smaller than the potential energy density, i.e. $\dot\phi^2 \ll V(\phi)$, and smallness of $\eta$ indicates $\ddot\phi \ll H\dot\phi$. These conditions simplify considerably the dynamical equations so that
\begin{equation}\label{friedmann_simplified}
    3 M_p^2 H^2 \simeq V(\phi), \qquad  3 H \dot\phi \simeq V'(\phi).
\end{equation}

The amount of inflation is measured by the e-fold parameter, defined as 
\begin{equation}\label{efold}
    N = \int_{t_\star}^{t_e} H dt = \int_{\phi_\star}^{\phi_e} \frac{H}{\dot\phi} d\phi,
\end{equation}
in which the subscript ``e" stands for the end of inflation, and the subscript of ``$*$" indicates the value of the parameter at the time where the pivot scale $k_\star = 0.05$ crosses the horizon. It is found that to solve the problems of the hot big-bang theory successfully, one needs to  require to have about $60-65$ e-folds of the expansion during the inflationary phase.

To estimate the validity of an inflationary model, it is required to compare the result of the model with observational data. The most important observable parameters are the amplitude of the scalar perturbations, the scalar spectral index, and the tensor-to-scalar ratio, which, in terms of the slow-roll parameters, can be written in the forms
\begin{eqnarray}\label{perturbation_parameters}
    A_s = \frac{1}{8\pi^2 M_p^2} \; \frac{H^2}{\epsilon}, \quad
    n_s - 1 = -6 \epsilon + 2 \eta, \quad 
    r = 16 \epsilon.
\end{eqnarray}
Inflation ends as the slow-roll parameter $\epsilon$ reaches unity, and the field at the end of inflation is obtained by solving the relation $\epsilon(\phi_e) = 1$. Then, using Eqs.\eqref{friedmann_simplified} and (\ref{efold}), the slow-roll parameters are estimated at the time of the horizon crossing, i.e. about 60-65 e-folds of expansion before the end of inflation. The result is compared with the observational data and the valid values of the free parameters are investigated. To examine the present model, we will consider the recent data release (DR6) from the ACT collaboration, as mentioned above.

\subsection{Reheating phase}
At the end of the inflationary phase, the universe is in the super-cooled state, empty of any standard particles. To recover the radiation phase, the universe needs to warm up and be filled with standard particles. This stage is known as the reheating phase. Reheating is a post-inflationary phase, during which, due to the coupling between the inflaton and other standard fields, the energy transforms from the inflaton to the standard model particles. The particles interact each other and consequently heat up the universe. Then, the universe could enter the radiation-dominant phase \cite{Albrecht:1982mp,Shtanov:1994ce,Kofman:1997yn,Bassett:2005xm,Rehagen:2015zma,Cook:2015vqa}. 
The reheating phase is parametrized by three parameters: the number of e-folds during the reheating phase $N_{re}$, the thermalization temperature $T_{re}$, and the equation of state $\omega_{re}$ \cite{Martin:2014nya,Rehagen:2015zma}. At the end of the reheating phase, the energy density can be expressed in terms of the reheating temperature as $\rho_{re} = \frac{\pi^2}{30} g_{\star re} T_{re}^4$. Then, by assuming a constant $\omega_{re}$ during the reheating phase, the reheating number of e-fold is obtained as 
\begin{equation}\label{Nre}
    {N_{re}} = -\frac{1}{3(1+\omega_{re})}\ln\left(\frac{\pi^2 g_{\star re}}{90 M_p^2 H_e^2}\right) - \frac{4}{3(1+\omega_{re})}\ln\left(T_{re}\right),
\end{equation}
where $g_{\star re}$ is the number of relativistic degrees of freedom in the thermal bath. \\
On the other hand, from the conservation of the comoving entropy from the reheating time to the present time, one can find the reheating temperature as
\begin{equation}\label{Tre1}
    T_{re} = \left( \frac{43}{11 g_{\star s, re}} \right)^{1/3} \; \frac{H_k \; T_0}{k_\star} \; e^{-(N_k + N_{re})},
\end{equation}
in which $H_k$ is the Hubble parameter at the horizon crossing, $N_k$ is the number of e-fold from the horizon crossing to the end of the inflationary phase, and $T_0$ is the current CMB temperature, $T_0 = 2.735 \; K = 2.35 \times 10^{-4} \; \text{eV}$. Additionally, the scale factor at the present time is taken to be $a_0 = 1$. Substituting the reheating e-fold \eqref{Nre} in Eq.\eqref{Tre1}, one arrives at \cite{Cook:2015vqa}
\begin{eqnarray}\label{Tre2}
    T_{re}  =  \left[ \left( \frac{11 g_{\star s, re}}{43} \right)^{1/3} \; 
    \left( \frac{90 M_p^2 H_e^2}{\pi^2 g_{\star re}} \right)^{\frac{1}{3(1+\omega_{re})}} \right. 
     \frac{k_\star}{H_k \; T_0} \; e^{N_k} \Bigg]^{\frac{3(1+\omega_{re})}{1-3\omega_{re}}}. 
\end{eqnarray}
The above relations provide a direct connection between the inflationary phase and the following reheating phase. Using the inflationary e-fold and values of the Hubble parameter obtained at the horizon crossing and at the end of inflation, one can estimate the values of the reheating temperature in terms of the reheating equation of state. 

The reheating temperatures span a wide range. There is a BBN temperature, $T_{BBN} \simeq 4 \; {\rm MeV}$ \cite{Kawasaki:1999na,Kawasaki:2000en,Hasegawa:2019jsa}, which is the lower bound for the reheating temperature. And we also have the GUT temperature as ($\simeq 10^{16} \; \rm GeV$), which is the upper limit for $T_{re}$. However, the considerations of 
PGWs 
can put further severe constraints \cite{Boyle:2005se,Watanabe:2006qe,Saikawa:2018rcs,Caprini:2018mtu,Figueroa:2019paj,Bernal:2019lpc,Bernal:2020ywq,Bhattacharya:2020lhc,Bhattacharya:2019bvk}. The PGWs evolve through the post-inflationary phase and contribute to the effective number of relativistic species. This contribution could be significant 
especially for a reheating phase with a stiff equation of state, i.e., $\omega > 1/3$. This contribution affects the reheating history and should be studied with care.

\subsection{PGW and the number of relativistic species}

Inflation not only solves the classical problems of the Big Bang model, but also explains the generation of primordial fluctuations. The scalar fluctuations that give rise to the large-scale structure have received lots of attention. In addition, inflation also foretells the creation of tensor perturbations:  PGWs. These PGWs can have a significant impact on the effective number of relativistic species, $\Delta N_{\rm eff}$, especially if there exists a non-standard phase after the inflation where the Eos ($\omega$) can be $\geq 1/3$. Since the contributions from $\Delta N_{\rm eff}$ place strict constraints on the reheating history and the feasibility of the inflationary model, it is important to take the evolution of the universe during this phase  into considerations. The frequency tail of the PGWs can be impacted by the presence of the stiff fluid  where $\omega \geq 1/3$. The extra energy injection by these species can alter the abundance of the light elements during the BBN. The recent observations of ACT, along with Planck, impose the server constraints on $\Delta N_{\rm eff} \leq 0.17$. Following \cite{Haque:2021dha,Maity:2024odg,Chakraborty:2023ocr}, we can use this bound to constrain the present-day energy density of the PGWs
\begin{equation}
     \int_{k_{\text{RH}}}^{k_{\text{end}}} \frac{dk}{k} \, \Omega_{\mathrm{GW}}^{(0)}(k) h^2 \leq \frac{7}{8} \left( \frac{4}{11} \right)^{4/3} \Omega_{\gamma}^{(0)} h^2 \, \Delta N_{\mathrm{eff}},
     \label{pgw1}
\end{equation}
where $\Omega_{\gamma}^{(0)} h^2 \approx 2.47 \times 10^{-5}$ denotes the current radiation energy density. Since there is a growth in the spectral energy density of the PGWs for the modes that enter the horizon during reheating ($k > k_{\rm RH}$). This becomes more significant for an epoch when $\omega \geq 1/3$. A simpler form of the equation \ref{pgw1} can be written after considering the form of $\Omega^{(0)}_{\rm GW}$ and $\Omega^{(0)}_{\rm R}$ for the radiation epoch
\begin{equation}
\Omega_R^{(0)} h^2 \, \frac{H_{\text{e}}^2 \; \mu(w_{\text{re}})}{12 \pi M_P^2}  \, \frac{(1 + 3 w_{\text{re}})^2 }{3 w_{\text{re}} - 1} 
 \left( \frac{k_{\text{e}}}{k_{\text{re}}} \right)^{\frac{6 w_{\text{re}} - 2}{1 + 3 w_{\text{re}}}}  \leq 5.61 \times 10^{-6} \, \Delta N_{\text{eff}},
\end{equation}
where $\mu_{RH}$ is a $\mathcal{O}(1)$ parameter which is defined as 
\begin{equation}
    \mu(\omega_{re}) = \big( 1 + 3\omega_{re} \big)^{4 \over 1+3\omega_{re}} \; 
      \Gamma^2\left( \frac{5+3\omega}{2 + 6\omega_{re}} \right).
\end{equation}
Once we obtain the $k_{\rm end}$ and $k_{\rm RH}$, it is rather straightforward to obtain a lower bound on the reheating temperature $T_{\rm re}$ following \cite{Haque:2025uri,Haque:2025uis,Haque:2021dha}
\begin{equation}
T_{\text{re}}  \geq  
\left[ 
  \frac{\Omega_R^{(0)} h^2}{5.61 \times 10^{-6} \, \Delta N_{\text{eff}}}
~ \frac{H_{\text{e}}^2 \; \mu(w_{\text{re}})}{12 \pi M_P^2}
~\frac{(1 + 3 w_{\text{re}})^2}{3 w_{\text{re}} - 1}
\right]^{\frac{3(1 + w_{\text{re}})}{4 (3 w_{\text{re}} - 1)}} \left( \frac{90 \, H_{\text{e}}^2 M_P^2}{\pi^2 g_{*\text{re}}}\right)^{\frac{1}{4}}\equiv T_{\text{re}}^{\text{GW}}.
\end{equation}

\section{Results \label{results}}

Beyond the Poincaré symmetry, supersymmetry is the last type of space-time symmetry permitted by the Coleman-Mandula no-go theorem~\cite{PhysRev.159.1251}. Following this, a generalized potential was derived in~\cite{Dimopoulos:2016zhy} and can be expressed as \cite{Yogesh:2024mpa}
\begin{equation}
    V(\phi) = V_0 \; \left( \frac{\phi^n}{\phi^n + m^n} \right)^q,
\end{equation}
where $V_0$ is a positive constant, $m$ is the mass, and $n$ and $q$ are real parameters. Also, it is assumed that $\phi > m$, otherwise the potential would be similar to the monomial inflation with $V(\phi) \propto \phi^{nq}$ \cite{Dimopoulos:2014boa}, which has been ruled out by observation. 

\begin{figure}[!ht]
    \centering
    \subfigure[$n=1$ \label{rns_n1q1}]{\includegraphics[width = 7cm]{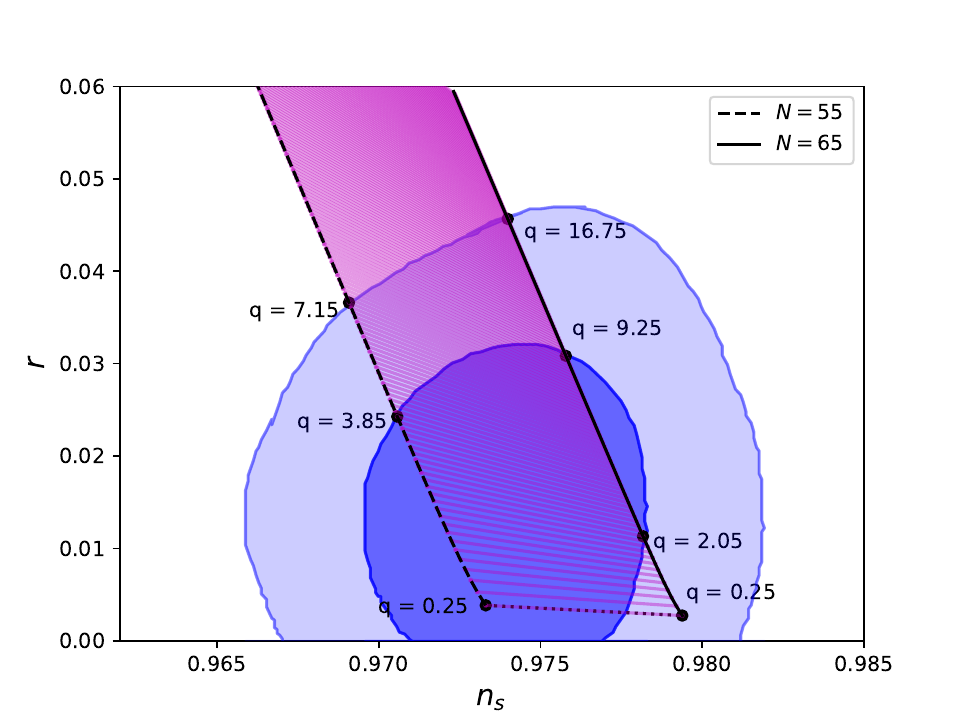}}
    \subfigure[$n=2$ \label{rns_n1q2}]{\includegraphics[width = 7cm]{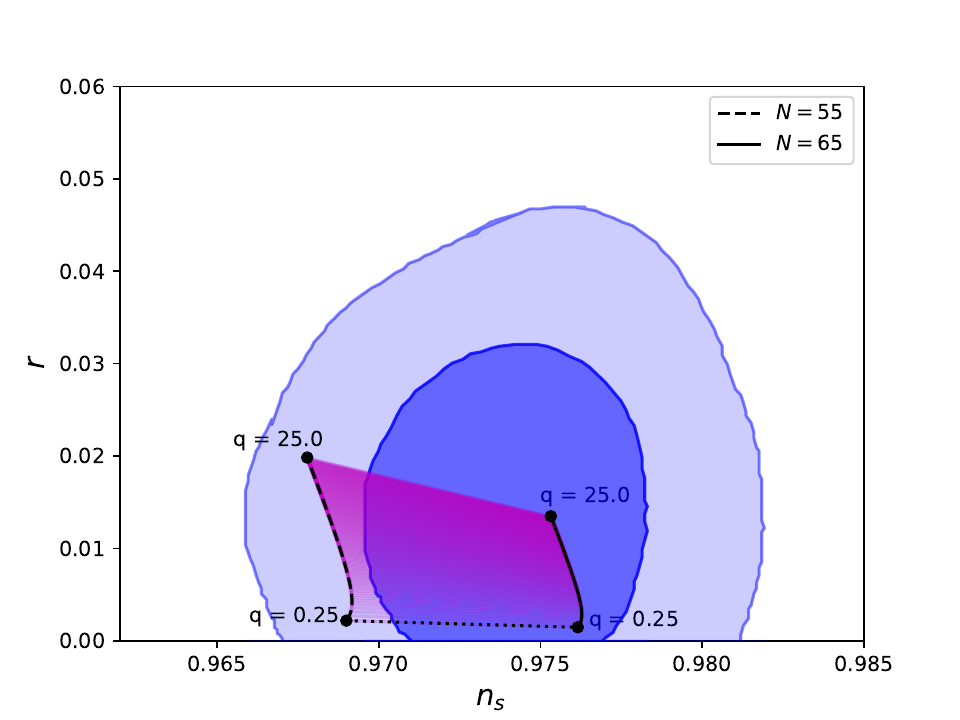}}
    \subfigure[$q=1$ \label{rns_n2q1}]{\includegraphics[width = 7cm]{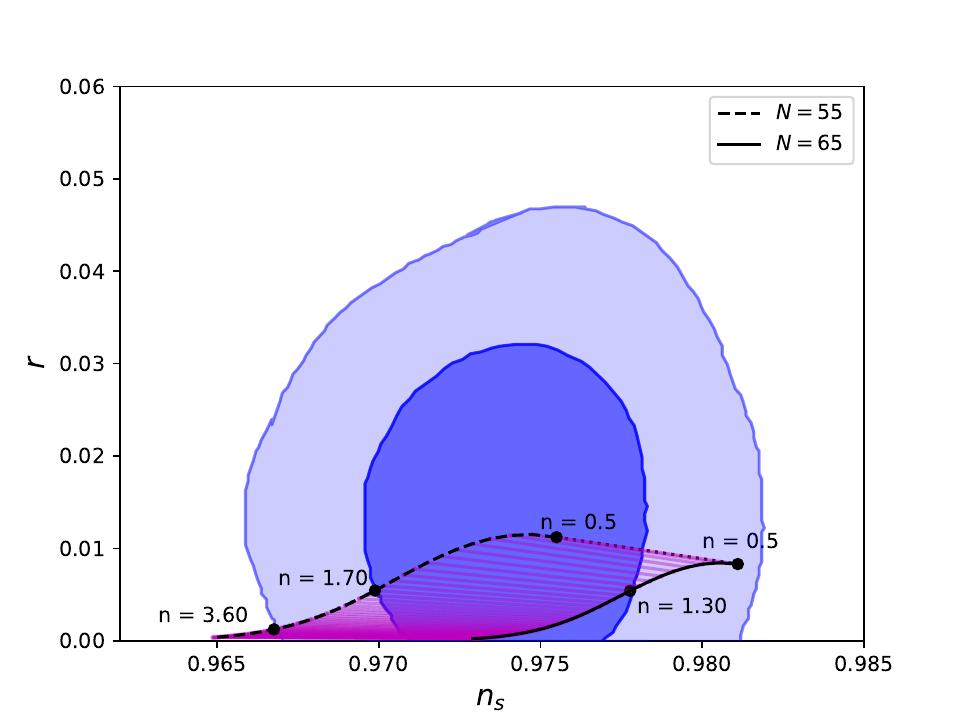}}
    \subfigure[$q=2$ \label{rns_n2q2}]{\includegraphics[width = 7cm]{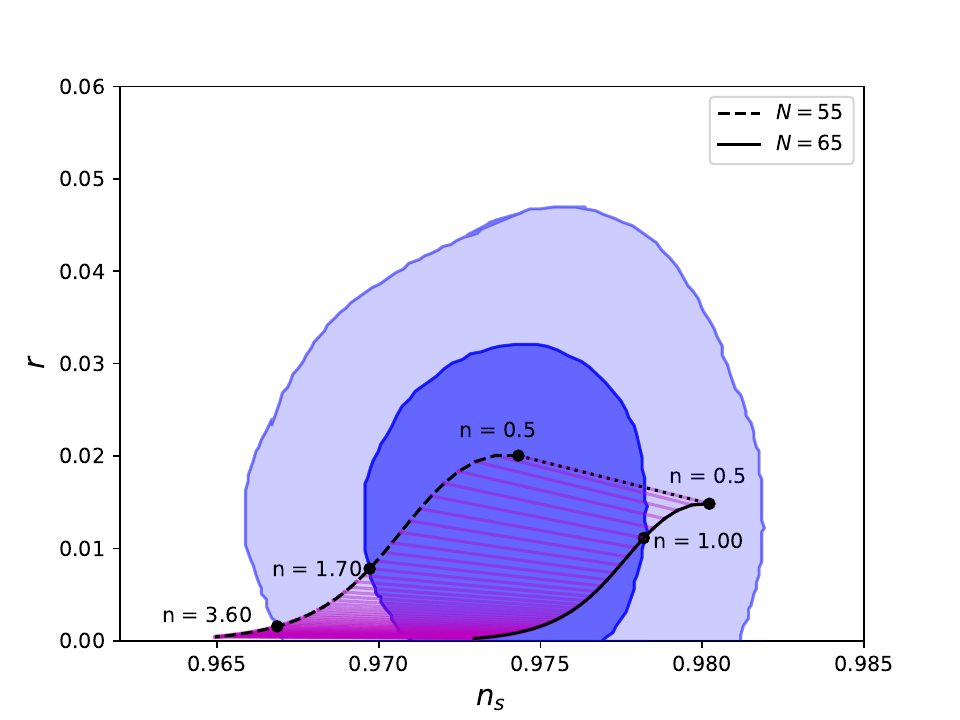}}
    \caption{The plots display the $r-n_s$ results of the model; compared with the ACT DR6, for different values of $n$ and $q$ parameters. The black solid and dashed lines are related to the number of e-folds $N=65$ and $55$, respectively. In the subfigures (a) and (b), the $n$ parameter is taken as a constant, while $q$ varies. However, in the subfigures (c) and (d), the $q$ parameter is taken as a constant, and the parameter $n$ varies. It is realized that the model can remain in the $1\sigma$ region and agree well with the ACT data. }
    \label{rns_nq}
\end{figure}

In this section, we discuss the results of the present model and compare them with the ACT DR6. 
Substituting the potential in the first slow-roll parameter \eqref{srp}, and solving the relation $\epsilon(\phi_e) = 1$, the scalar field at the end of inflation is obtained. Then, using this result, we solve the e-fold integral and find the value of the field at the horizon crossing time, i.e. $\phi_\star$, so that  $N \gtrsim 60-65$  is generated. Utilizing the resulting $\phi_\star$, we can obtain the scalar spectral index and the tensor-to-scalar ratio  at the horizon crossing. The resulting $n_s$ and $r$ do not depend on the constant $V_0$. For comparison with the observational data, the results are illustrated in Fig.\ref{rns_nq} for different values of $n$ and $q$, with the number of e-folds fixed at $N=55$ and $N=65$. In this plot, the solid and dashed black lines are related to the number of e-folds $N = 55$ and $N = 65$, respectively. 
\begin{itemize}
    \item Fig.\ref{rns_n1q1} shows the $r-n_s$ diagram for $n=1$, and the parameter $q$ is taken as the varying parameter, varies from $0.25$ to $25$. By increasing the parameter $q$, the scalar spectral index decreases; however, the scalar spectral index increases. Although for $N=65$, the result initially remains within $2\sigma$, it enters the $1\sigma$ region at $q = 2.05$ and remains in this region until $q=9.25$. By increasing the $q$ parameter, the curves enter the $2\sigma$ region, which occurs at $q = 3.85$ and $9.25$ for $ N = 55$ and $65$, respectively. Then, they will be out of both the $1\sigma$ and $2\sigma$ regions for all values of the number of e-folds $N$. The curve leaves the observational range for $ q > 7.15 \; (16.75)$ for $N = 55 \; (65)$. Therefore, for the number of e-folds $N=55 \; (65)$, the results are within $1\sigma$ for a specific range of the $q$ parameter, i.e. $q \in [0.25, 3.85] \; ([2.05, 9.75])$, and agree well with the observational data. It should be noted that for higher values of $N$, the acceptable range of $q$ becomes wider \footnote{In~\cite{Dimopoulos:2016zhy}, the authors considered $n$ and $q$ from 1 to 4; here we extend these values to a wider range.}.  
    
    \item The $r-n_s$ diagram in Fig.\ref{rns_n1q2} is plotted for $n=2$ with varying $q$ parameter in the range $q \in [0.25,25]$. In contrast to the previous case, it is realized that for $N=55$, the result is out of $1\sigma$ region and it stays in $2\sigma$ region. The curve shows a tendency to leave the $2\sigma$ region for higher values of $q$. However, for $N=65$, the result always remains within the $1\sigma$ region for a wide range of the $q$ parameter, indicating consistency with observational data. 
    
    \item In Fig.\ref{rns_n2q1}, the $r-n_s$ diagram is provided for $q=1$, and $n$  varies from $0.5$ to $6$. By increasing the $n$ parameter, both the scalar spectral index and the tensor-to-scalar ratio decrease. For $N=65$, the result is in the $2\sigma$ region; however, it then enters the $1\sigma$ region at $n = 1.30$ and remains in this region for higher values of $n$. On the other hand, for $N=55$, the result initially falls within the $1\sigma$ region, and as the $n$ parameter increases, it enters the $2\sigma$ region at $q = 1.70$, eventually leaving the region at $q = 3.60$. 
    
    \item The $r-n_s$ diagram for $q=2$ and varying $n$ is plotted in Fig.\ref{rns_n1q2}. The situation is similar to  the previous case. The only difference is that, in this case, for each value of the $n$ parameter, the tensor-to-scalar $r$ is larger, and the scalar spectral index $n_s$ is slightly smaller.
\end{itemize}

In general, from the plots one can see that the resulting $r-n_s$ can remain within the $1\sigma$ region of the observational data. 
\begin{table}[h]
    \centering
    \begin{tabular}{c|c
                    >{\hspace{0.5em}}c<{\hspace{0.5em}}  
                    >{\hspace{0.5em}}c<{\hspace{0.5em}}  
                    >{\hspace{0.5em}}c<{\hspace{0.5em}}  
                    >{\hspace{0.5em}}c<{\hspace{0.5em}}  
                    >{\hspace{0.5em}}c<{\hspace{0.5em}}  
                    c}                                  
         & \ $q$ \ & $n_s$ & $r$ & $H_s$ & $H_e$  & $\phi_s$ & $\phi_e$  \\[0.3em]
    \hline \\[-0.5em]
       \multirow{4}{*}{$n=1$}  & $1$ & $0.9769$ & $0.0077$ & $8.96 \times 10^{-6}$ &  $5.57 \times 10^{-6}$ & $5.189$ & $0.478$ \\ [0.3em]

         & $2$ & $0.9763$ & $0.0123$ & $1.13 \times 10^{-5}$ &  $5.75 \times 10^{-6}$ & $6.647$ & $0.790$ \\[0.3em]
       
         & $3$ & $0.9758$ & $0.0162$ & $1.29 \times 10^{-5}$ &  $5.67 \times 10^{-6}$ & $7.672$ & $1.039$ \\[0.3em]
       
         & $4$ & $0.9753$ & $0.0197$ & $1.42 \times 10^{-5}$ &  $5.52 \times 10^{-6}$ & $8.489$ & $1.254$ \\[0.3em]
       \hline \\[-0.5em]
       \multirow{4}{*}{$n=2$}  & $1$ & $0.9742$ & $0.0031$ & $5.73 \times 10^{-6}$ &  $3.34 \times 10^{-6}$ & $4.670$ & $0.83$ \\ [0.3em]

         & $2$ & $0.9741$ & $0.0044$ & $6.78 \times 10^{-6}$ &  $3.09 \times 10^{-6}$ & $5.477$ & $1.181$ \\[0.3em]
       
         & $3$ & $0.9741$ & $0.0054$ & $7.48 \times 10^{-6}$ &  $2.83 \times 10^{-6}$ & $6.079$ & $1.414$ \\[0.3em]
       
         & $4$ & $0.9740$ & $0.0062$ & $8.02 \times 10^{-6}$ &  $2.60 \times 10^{-6}$ & $6.544$ & $1.595$ \\[0.3em]
       \hline \\[-0.5em]
       \multirow{4}{*}{$n=3$}  & $1$ & $0.9728$ & $0.0013$ & $3.80 \times 10^{-6}$ &  $2.75 \times 10^{-6}$ & $3.865$ & $1.023$ \\[0.3em] 

         & $2$ & $0.9728$ & $0.0018$ & $4.34 \times 10^{-6}$ &  $3.03 \times 10^{-6}$ & $4.452$ & $1.308$ \\[0.3em]
       
         & $3$ & $0.9728$ & $0.0021$ & $4.70 \times 10^{-6}$ &  $3.19 \times 10^{-6}$ & $4.834$ & $1.486$ \\[0.3em]
       
         & $4$ & $0.9728$ & $0.0023$ & $4.97 \times 10^{-6}$ &  $3.30 \times 10^{-6}$ & $5.124$ & $1.618$ \\[0.3em]
       \hline \\[-0.5em]
        \multirow{4}{*}{$n=4$} & $1$ & $0.9718$ & $0.0007$ & $2.71 \times 10^{-6}$ &  $2.12 \times 10^{-6}$ & $3.347$ & $1.113$ \\ [0.3em]

         & $2$ & $0.9719$ & $0.0008$ & $3.03 \times 10^{-6}$ &  $2.33 \times 10^{-6}$ & $3.762$ & $1.339$ \\[0.3em]
       
         & $3$ & $0.9719$ & $0.0010$ & $3.24 \times 10^{-6}$ &  $2.45 \times 10^{-6}$ & $4.028$ & $1.476$ \\[0.3em]
       
         & $4$ & $0.9719$ & $0.0011$ & $3.40 \times 10^{-6}$ &  $2.53 \times 10^{-6}$ & $4.227$ & $1.576$ \\[0.3em]
       \hline
    \end{tabular}
    \caption{The table shows the numerical results of the model for different values of the free parameters $n$ and $q$, for $N = 60$. The results remain within $1\sigma$ region and demonstrate a good agreement with the ACT data. }
    \label{table_rns}
\end{table}
Table.\ref{table_rns} provides the numerical results of the model for different values of the free parameters $n$ and $q$ and for the inflation e-folds $N = 60$. As it was expected, the resulting $r-n_s$ remains in the $1\sigma$ region of data. It is also demonstrated from the table that, for higher fixed values of the $n$ parameter, the scalar spectral index and the tensor-to-scalar ratio remain almost the same by changing the $q$ parameter. While increasing $q$ amplifies the differences in both the field values ($\Delta \phi = \phi_\star-\phi_e$) and Hubble parameters ($H_\star - H_e$) between horizon crossing and inflation's end, these differences are suppressed at higher $n$ values. The energy scale of inflation is around $\simeq 10^{-3} M_p$, which can be found in the table above.

\begin{figure}[!h]
    \centering
    \subfigure[$n=1$ and $q=1$ \label{Tre_n1q1}]{\includegraphics[width=6cm]{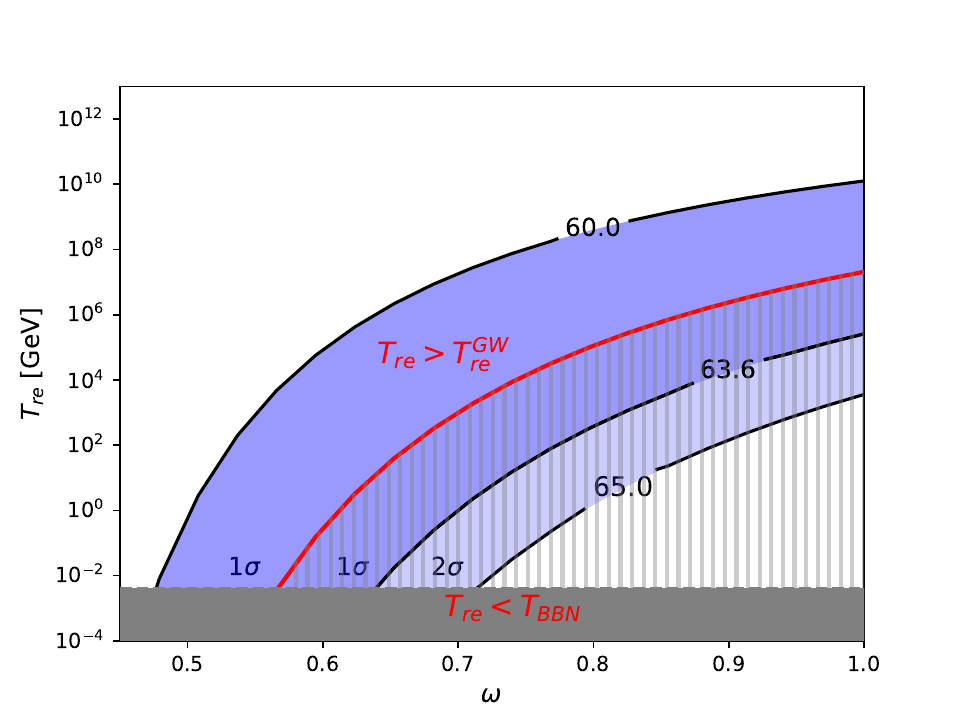}}
    \subfigure[$n=1$ and $q=2$ \label{Tre_n1q2}]{\includegraphics[width=6cm]{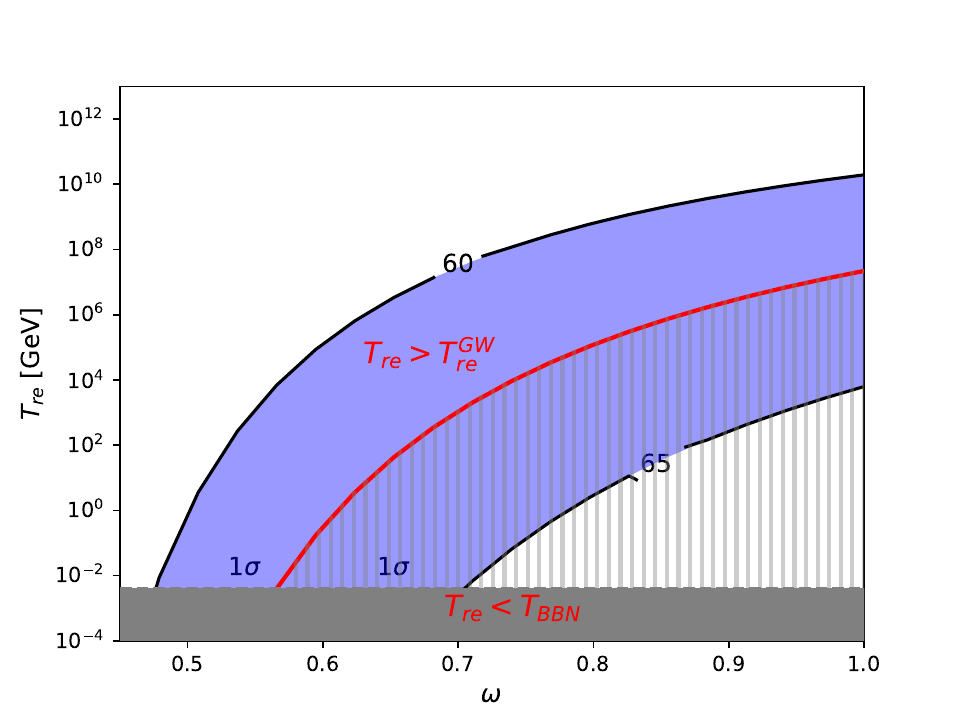}}
    \subfigure[$n=2$ and $q=1$ \label{Tre_n2q1}]{\includegraphics[width=6cm]{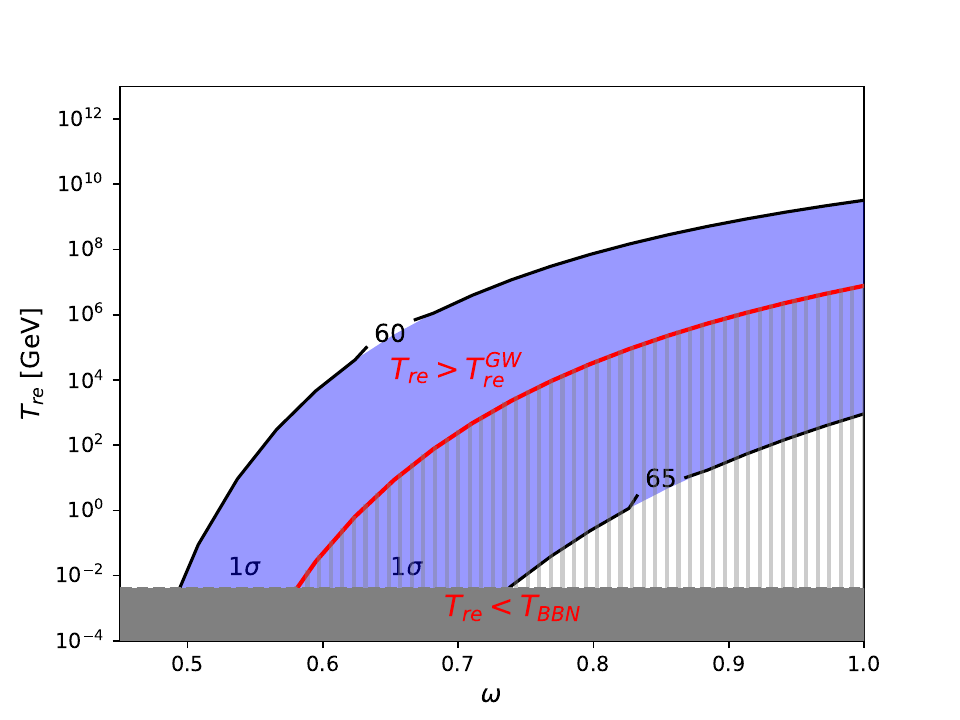}}
    \subfigure[$n=2$ and $q=2$ \label{Tre_n2q2}]{\includegraphics[width=6cm]{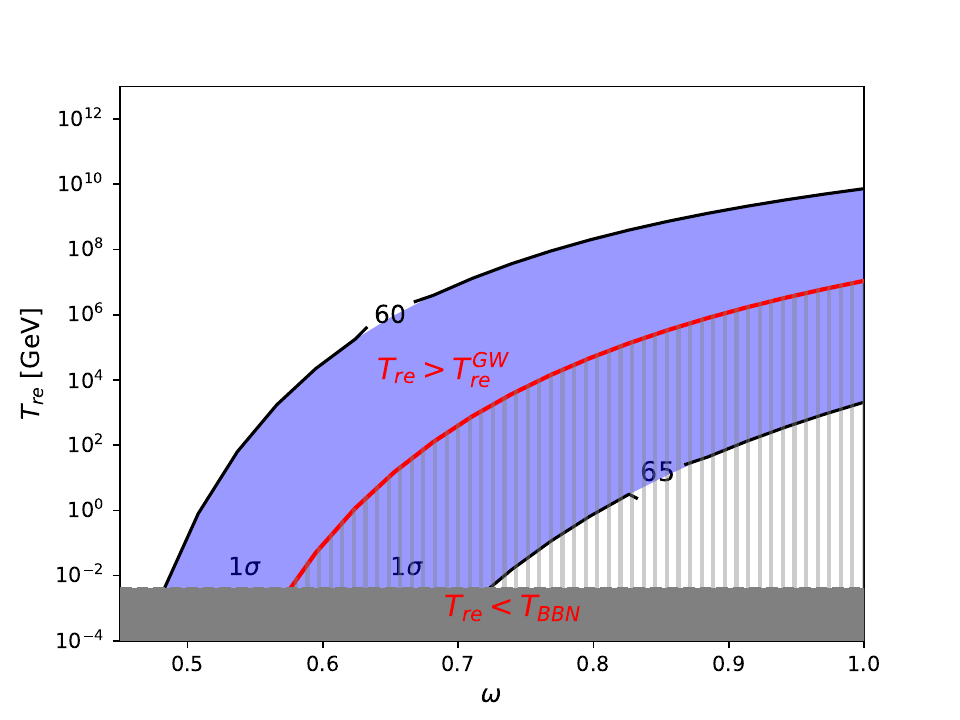}}
    \caption{The reheating temperature versus the reheating effective equation of state parameter is plotted for different choices of the $n$ and $q$ parameters. The gray region is the BBN temperature bound, and the solid red line is the $T_{re}^{GW}$ bound. The blue-colored area represents the region where the model falls within the $1\sigma$ region, and the light blue-colored area corresponds to the $2\sigma$ region. The shaded area is the region where the reheating temperature falls below the $T_{re}^{GW}$ bound. The figure shows that the $T_{re}^{GW}$ bound is efficient for $\omega_{re} > 0.58$. To satisfy the new reheating bound and also to put the result of the model in the $1\sigma$ region of observational data, the total number of e-folds during the inflationary phase should be $N_k \lesssim 62$. }
    \label{Tre}
\end{figure}

After the inflationary phase, the universe enters the reheating phase, during which particles are produced as a result of the energy transformed from the inflaton. The particles interact each other and the universe gets warmed up. To parameterize the reheating phase, $T_{re}$, $\omega_{re}$, and $N_{re}$ is our main parameters. The reheating temperature is obtained as Eq.\eqref{Tre2}, which is related to the number of e-folds of the inflationary phase and also the Hubble parameter at the time of horizon crossing and the time of the end of the inflationary phase. The dependency of the reheating temperature on these parameters implies a connection between the inflationary phase and the following reheating phase. In general, there is a wide range for the reheating temperature, which varies from the end of inflation ($\simeq 10^{16} \; \rm Gev$ ) to the BBN epoch $4 \; \rm MeV$. However, reheating could have a significant influence on the PGWs and may lead to an increase in their amplitude. Due to this enhancement, the total energy density of the PGWs increases, and the effective number of the relativistic species is affected. Therefore, the observational data for $\Delta N_{\rm eff}$ could also provide a lower limit on the reheating temperature. The enhancement of the PGWs would be more efficient for the stiff equation of state, i.e. $\omega_{re} > 1/3$. Therefore, we will consider only the stiff equation of state.  

Fig.\ref{Tre} displays the reheating temperature versus the equation of state $\omega_{re}$ for different choices of the $n$ and $q$ parameters. It is also assumed that the number of e-folds during inflation should be around $60-65$.
The behavior of the reheating temperature is illustrated in Fig.\ref{Tre}. Here, the solid black line represents the points where the number of e-folds $N_k$ is constant. In addition, the solid red line represents the temperature constraint resulting from $\Delta N_{\rm eff}$, which is bounded by $\Delta N_{\rm eff} \leq 0.17$ at $95\%$CL resulted from the combination of ACT and Planck. The figure describes the reheating temperature versus the equation of state  $\omega_{re}$ for the number of e-folds in the range $N_k=[60, 65]$ and for different choices of the parameters $n$ and $q$, so we have the following:
\begin{itemize}
    \item The blue color region describes the reheating temperature for a range of the number of e-folds $N_k$, where the results of inflation agree with the $1\sigma$ region of the ACT data. Similarly, the light blue area indicates the same result for a range of $N_k$ where the results of inflation match in the $2\sigma$ region of ACT data. 
    
    \item The magenta color area indicates the BBN constraint, i.e. $4 \; \text{MeV}$, so that the resulting reheating temperature must be larger than this constraint.
    
    \item The solid red color is the $T_{re}^{GW}$ constraint. The line divides the space of the plot into two areas:  $T_{re} > T_{re}^{GW}$ and  $T_{re} < T_{re}^{GW}$. Since the temperature $T_{re}^{GW}$ is the lower limit for the reheating temperature, only the upper area is acceptable. 
    
    \item The $T_{re}^{GW}$ limit implies that this upper bound is more efficient for higher values of the equation of the state parameter. For $\omega_{re} < 0.6$, the temperature $T_{re}^{GW}$ falls below the BBN temperature; however, by increasing $\omega_{re}$, the temperature $T_{re}^{GW}$ increases as well, whereby a stronger constraint on the reheating temperature is obtained.
\end{itemize}

From the figure, it is realized that the area between the red line and the black line with $N_k = 60$ is the acceptable range. Therefore, the total number of e-folds during the inflationary phase should be almost in the range of $60$ to $62$ to achieve both a successful inflationary phase and a consistent reheating temperature. Considering the values of the equation of state parameters, it is evident that increasing $\omega_{re}$ allows for higher values of the reheating temperature. On the other hand, by increasing the inflation e-fold $N_k$, the reheating temperature decreases and it falls below the $T_{re}^{GW}$ line. Therefore, although the result of the inflationary phase agrees with the $1\sigma$ of the observational data, the $\Delta N_{\rm eff}$ bound will be violated.

\begin{figure}
    \centering
    \includegraphics[width=0.5\linewidth]{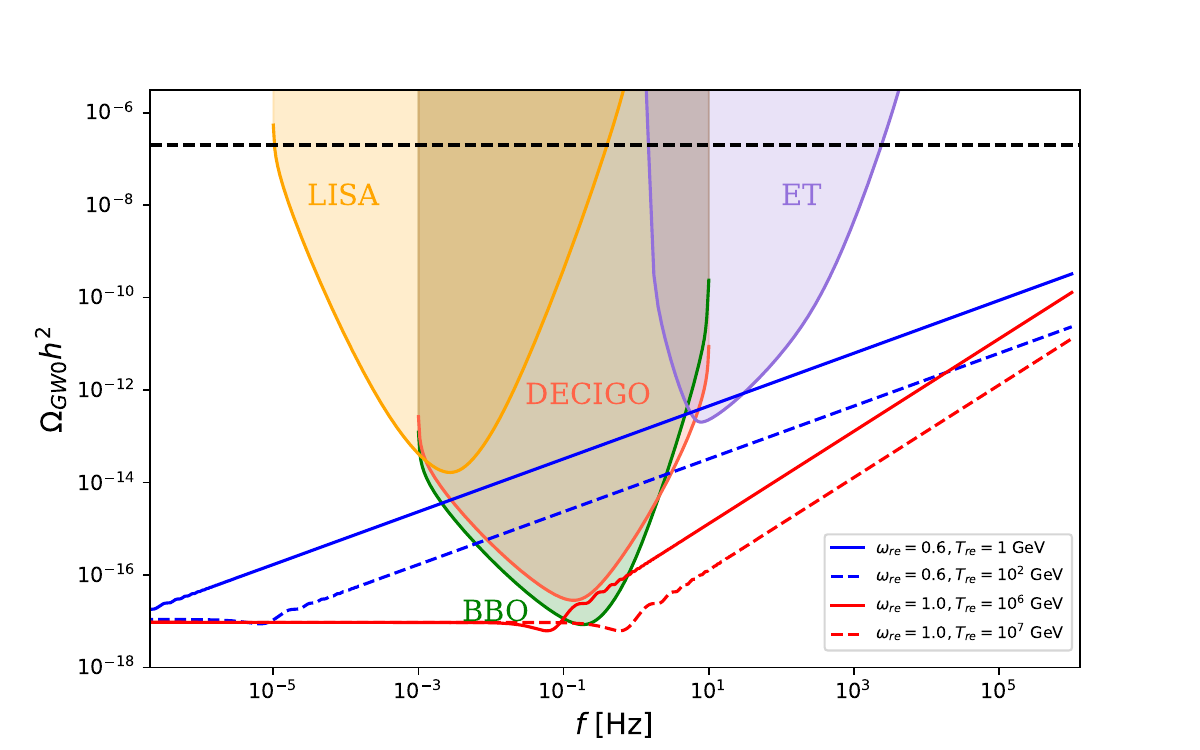}
    \caption{The plot displays the energy spectrum of the gravitational waves versus the frequency for different values of the effective equation of state $\omega_{re}$. The reheating temperature for each case of $\omega_{re}$ is chosen in the accepted range as shown in Fig.\ref{Tre}. Additionally, the free parameters are set to $n=q=1$.  }
    \label{gw}
\end{figure}

This new limit on the reheating temperature also affects the amplitude of the PGWs. Fig.\ref{gw} shows the energy spectrum of the gravitational waves for different values of the reheating effective equation of states. Based on the above result, from Fig.\ref{Tre} it is found that the allowed range of reheating temperature for each value of $\omega_{re}$ is different. Therefore, we only select the values of the reheating temperature that satisfy the aforementioned bound. A smaller value of the reheating temperature means that the universe stays longer in the reheating phase. Consequently, more modes re-enter the Hubble horizon, and the effect of reheating appears as more amplification on the amplitude of the energy density of the gravitational waves. Therefore, we have a higher magnitude by decreasing the reheating temperature for each value of $\omega_{re}$. Moreover, increasing the effective equation of state of the reheating results in an enhancement in the magnitude of the energy spectrum of gravitational waves.   

\section{Conclusion \label{conclusion}}

The scenario of inflation in the standard gravity model has been considered, where the inflaton has a minimal coupling to gravity. Motivated by supersymmetry, the potential of the inflaton is considered to be the PLP potential. Fitting the model with the ACT DR6, it was found that the model can remain well within the $1\sigma$ region of the ACT data for a wide range of the free parameters of the model, displaying the perfect agreement between the model and the observational data.

In considering the reheating phase, we focused on the reheating temperature as one of the key parameters that determines the dynamics of the reheating process. In addition to the BBN temperature, which plays the role of a lower bound for the reheating temperature, there is another bound that originates from the effective number of relativistic species. The produced PGWs remain scale-invariant as they exit the Hubble horizon in the inflationary phase. Then, after the end of inflation, the modes that re-enter the horizon will be affected in the reheating phase. Consequently, they contribute to the effective number of relativistic species, which has been constrained by CMB data to $\Delta N_{\rm eff} \leq 0.17$ as a result of a measurement combining ACT and Planck data. Our considerations presented in this paper showed that the new bound, i.e., $T_{re}^{GW}$, would be more efficient for stiff $\omega_{re}$, or more precisely for $\omega_{re} > 0.58$. It was realized that for the model to stay in $1\sigma$ region of the ACT data and simultaneously satisfy the $T_{re}^{GW}$ bound of the reheating temperature, the total number of e-folds during the reheating phase should be $N_k \lesssim 62$. Considering the PGWs, we found that a smaller reheating temperature results in a higher enhancement of the magnitude of the energy spectrum of the gravitational waves. On the other hand, a smaller value of the reheating effective equation of state leads to a lower magnitude for the energy spectrum. In addition, there is a good chance for future detections for the larger $\omega_{re}$. Also, for smaller $T_{re}$, the curves cross the observatory range for a broader range of frequencies, leading to a higher chance of detections.

\acknowledgments

A.W. is partly supported by the US NSF grant, PHY-2308845.

\bibliographystyle{apsrev4-1}
\bibliography{plpACTbib}

\end{document}